\documentclass[twocolumn,showpacs,amsmath,amssymb,pra]{revtex4}

\usepackage{graphicx,bm,bbm,amsfonts}

\newcommand{\bra}[1]{\langle #1|}
\newcommand{\ket}[1]{|#1\rangle}

\newcommand{\tr}[2]{{\,\rm tr_{#1}}{\lbrack #2 \rbrack}\,}
\def\ii{{\rm i}}

\begin{document}

\title{Optimal two-qubit gate for generation of random bipartite entanglement}

\author{Marko \v Znidari\v c}

\affiliation{Department of Physics, Faculty of Mathematics and Physics, University of Ljubljana, SI-1000 Ljubljana, Slovenia}

\begin{abstract}
We numerically study protocols consisting of repeated applications of two qubit gates used for generating random pure states. A necessary number of steps needed in order to generate states displaying bipartite entanglement typical of random states is considered. Numerics indicates that for a generic two qubit entangling gate the decay of purity is exponential with the decay time scaling as $\sim n$, implying that of order $\sim n^2$ steps are needed to reach random bipartite entanglement. We also numerically identify the optimal two qubit gate for which the convergence is the fastest. Perhaps surprisingly, applying the same good two qubit gate in addition to a random single qubit rotations at each step leads to a faster generation of entanglement than applying a random two qubit transformation at each step.   
\end{abstract}
\maketitle

\section{Introduction}
Entanglement is one of the resources which can be used to perform tasks not possible by classical means. Considerable effort has been put into understanding entanglement properties of various states. While entanglement of certain classes of states, for instance of random states, is well understood, quantifying entanglement of a general quantum state is a rather difficult task, for a review see, {\em e.g.}~\cite{Ent_review}. In the present paper we are going to study optimal protocols for producing random states from initial product states. By random states we mean pure quantum states which are eigenvectors of random unitary matrices distributed according to the unitarily invariant Haar measure. Random states have been the subject of considerable interest in the past. Entropy and purity of a subsystem have been studied in~\cite{Lubkin:78,Lloyd:88,Page:93,Ruiz:95,Sen:96,Zyczkowski:01,Kendon:02}. Distribution of the inverse of purity has been considered in~\cite{Facchi:06}, the distribution of G concurrence (geometric mean of Schmidt coefficients) has been derived in~\cite{Cappellini:06}, the distribution of purity in~\cite{Olivier:06} and the average values of Schmidt coefficients in~\cite{JPA}. It has been shown that random states reproduce certain statistical properties of eigenstates of quantum systems whose classical counterparts are chaotic very well. Even though a particular chaotic system has a well defined Hamiltonian a statistical description with random Hamiltonian, the so-called random matrix theory~\cite{RMT}, has been very successful. 

One can ask if random states are also relevant for quantum information theory? Note that if description in terms of random states is applicable this usually simplifies the analysis~\cite{Hayden:06}. Random states are a directly needed resource in certain procedures like quantum dense coding~\cite{Harrow:04} or remote state preparation~\cite{Bennett:05,Hayden:04}. One area at which quantum physics can do better than classical is at computation, for instance, simulating quantum systems. If a state during quantum computation is not sufficiently entangled, efficient classical simulation is possible~\cite{Jozsa:03,Vidal:03}. Becouse random states are almost maximally entangled it is reasonable to expect that such random states will naturally occur during sufficiently complex quantum computation. In fact, we know that the evolution with chaotic systems will produce states whose statistical properties are well reproduced by random states. Therefore, quantum simulation of chaotic system is a likely candidate for problems which can be efficiently solved by quantum computer but not by classical. For examples of such algorithms see, {\em e.g.}~\cite{Benenti:01,Levi:03}. Note that sufficient entanglement is necessary but not sufficient for a quantum speed-up. An example are for instance quantum integrable systems which can also efficiently produce a lot of entanglement~\cite{Calabrese:05,Chiara:06} however, efficient classical simulation seems to be possible~\cite{MPS}. 

An important practical question is how to produce random states efficiently? By efficiently we mean in a number of steps that grows only polynomially with the number of qubits $n$. A way to achieve this is already suggested by chaotic dynamics. Evolution with chaotic dynamics will produce states which are generic, i.e., random, from almost any initial state. It is well established that bipartite entanglement as measured by purity or von Neumann entropy increases linearly with time for chaotic systems and that the asymptotic saturation is reached after time which grows linearly with the logarithm of the Hilbert space dimension~\cite{IS}. Typically, such chaotic Hamiltonian can be written as a sum of $n$ two qubit terms, and becouse entanglement saturates after time $\propto n$, this means that of order $\sim n^2$ two qubit gates are needed to generate a random state - at least as far as bipartite entanglement is concerned. Our results will confirm these expectations. One should be aware that to produce an arbitrary unitary transformation, and therefore a truly random state, an exponential number of two qubit gates is needed in general. However, if our criteria is just to reproduce bipartite entanglement of a typical random state, which is the case in the present paper, only polynomial number of gates is needed. The procedure to generate random states is therefore the following: generate some pseudo-random sequence of gates, thereby producing the so-called pseudo unitary operator, applying it to an arbitrary separable initial state. After sufficient number of steps, i.e., applications of random gates, we will end up in a random state. Such random protocol has been numerically studied in~\cite{Emerson:03,Weinstein:05} while a convergence to uniform Haar measure has been discussed in~\cite{Emerson:05}. The amount of interference produced by such protocols has been considered in~\cite{Braun:06}. Recently the question of how many gates do we need to obtain convergence has been attacked by analytical tools of Markovian chains~\cite{Plenio} and a lower bound on the number of steps needed has been proved. In the present paper we are going to numerically calculate the exact convergence times for small chains as well as find a two qubit gate which will result in the fastest generation of bipartite entanglement.

\section{Protocol}
\label{sec:protocol}
Let us denote a state at time $t$ by $\ket{\psi(t)}$. The protocol for generation of random entanglement consists of application of two qubit transformation at each time step
\begin{equation}
\ket{\psi(t+1)}=U_{ij}(t)\ket{\psi(t)},
\label{eq:evol}
\end{equation}
with $U_{ij}(t)$ being a two qubit gate acting on $i$-th and $j$-th qubits out of total $n$ qubits. At each time step the pair of qubits $i$ and $j$ on which a gate acts will be drawn independently. We will consider three cases (couplings): (i) $i$ and $j$ are chosen randomly, that is gate can act on an arbitrarily separated qubits; (ii) $i$ and $j$ are neighboring qubits, that is a par $(i,i+1)$ or $(i+1,i)$, and we take periodic boundary conditions, meaning that the first and the last qubits can also be coupled. Such coupling will be abbreviated nnPBC. (iii) similarly as in case (ii) we allow only nearest neighbor gates but with open boundary conditions, that is we do not allow a gate between the first and the last qubit. This will be abbreviated nnOBC. For the unitary matrix $U_{ij}$ we choose it to be a product of two independent single qubit unitaries $V_i$ and $V_j$ and two qubit gate $W_{ij}$,
\begin{equation}
U_{ij}(t)=V_i(t) V_j(t) W_{ij}.
\label{eq:U}
\end{equation}    
Two qubit gate will be the same for all steps whereas single qubit unitaries $V_i$ and $V_j$ will be chosen according to the U(2) invariant Haar measure at each time step and for each qubit independently, that is they are from CUE(2) ensemble. Motivation for such a protocol is that from the experimental perspective two qubit gates are difficult to make and therefore we always apply the same two qubit gate, whereas single qubit transformations are relatively simpler and can be changed at each step. Without sacrificing generality we will always choose the initial state to be a separable $\ket{\psi(0)}=\ket{00\ldots0}$. All the statements about the convergence times in the paper thus pertain only to separable initial states. Choosing an entangled initial state can presumably lead to a faster convergence.

The goal of entanglement generating protocol is to produce states whose entanglement is as close as possible to that of random pure states, i.e., states drawn according to the invariant Haar measure on $n$ qubits. Bipartite entanglement of a pure state is completely determined by its Schmidt coefficients $\mu_i$,
\begin{equation}
\ket{\psi}=\sum_i \mu_i \ket{\psi_i}_A \otimes \ket{\varphi_i}_B,
\end{equation}
where $\ket{\psi_i}_A$ and $\ket{\varphi_i}_B$ are orthogonal and we assume $\mu_i$ are listed in nonincreasing order. Stating all Schmidt coefficients completely characterizes bipartite entanglement~\cite{Vidal:00}. As a measure of closeness of our state $\ket{\psi(t)}$ to a random state, we could for instance compare average values of Schmidt coefficients. For random pure states the average value of $i$-th largest Schmidt coefficient $\mu_i$ has been calculated in~\cite{JPA}. In the case of a symmetric bipartite cut to first $n/2$ and last $n/2$ qubits it is given in an implicit form by
\begin{eqnarray}
\mu_i=\frac{2\cos{(\phi_i)}}{\sqrt{N}},\qquad \nonumber \\
\frac{(i+\frac{1}{2})\pi }{2 N}=\phi_i-\frac{1}{2}\sin{(2\phi_i)},
\label{eq:mui}
\end{eqnarray}
where $N=2^{n/2}$ is the dimension of subspaces. For analytical treatment though, Schmidt coefficients of $\ket{\psi(t)}$ are not the simplest quantity to calculate. Therefore, in most of the paper we will rather use purity as a measure of entanglement. Purity $I(t)$ is simply a sum of $\mu_i^4$, or in terms of reduced density matrix
\begin{equation}
I(t)=\tr{A}{\rho^2_A(t)},\qquad \rho_A(t)=\tr{B}{\ket{\psi(t)}\bra{\psi(t)}}.
\label{eq:I}
\end{equation}
Here subscript $A$ and $B$ denote subspaces of first and the last $n/2$ qubits, respectively. For random states purity has been calculated~\cite{Lubkin:78,Zyczkowski:01} and is $I(\infty)\asymp 2/N$. In the last part of the paper we will also briefly mention results for the von Neumann entropy,
\begin{equation} 
S(t)=-\tr{A}{\rho_A(t)\log_{2}{\rho_A(t)}}.
\label{eq:S}
\end{equation}
Value of $S$ for random states is known~\cite{Lloyd:88,Page:93} and is $S(\infty)\asymp n/2-1/\log_{\rm e}{4}$. Whenever we speak about purity or von Neumann entropy we will have in mind their approach to asymptotic values, i.e., $I(t)-I(\infty)$ and $S(\infty)-S(t)$.

The goal of the paper is to analyze how fast the purity decays to its asymptotic random state value, in particular, how the decay time scales with $n$. In addition we are going to find a two qubit gate $W$ (\ref{eq:U}) for which the (average) purity $I(t)$ will decay the fastest. As we will see, this also means that the convergence of $\mu_i$ or $S(t)$ is the fastest. In principle, to find the optimal gate, we would have to check the whole 15 parameter space of two qubit gates. However, each two qubit gate $W$ can be decomposed as~\cite{Khaneja:01,Kraus:01}
\begin{eqnarray}
W_{ij}=A_i \otimes B_j\,\, w(a_x,a_y,a_z)\,\, A'_i \otimes B'_j,\qquad \\
w(a_x,a_y,a_z)=\exp{\left(\ii \frac{\pi}{4}[a_x \sigma_i^x \sigma_j^x+a_y \sigma_i^y \sigma_j^y+a_z \sigma_i^z \sigma_j^z] \right)},\nonumber
\label{eq:w}
\end{eqnarray}
where $A,A',B$ and $B'$ are single qubit unitaries and $\sigma^{x,y,z}$ are standard Pauli matrices. Two qubit transformation $w(a_x,a_y,a_z)$ which is parameterized by three parameters is called a canonical form of $W$. In our protocol (\ref{eq:U}) single qubit gates $V$ are random unitaries therefore, as far as entanglement is concerned, we only have to consider two qubit gates in its canonical form. Instead of a 15 parameter space we only have to find the optimal gate among a 3 parameter set $w$. Furthermore, $w(a_x,a_y,a_z)$ has certain symmetries. For instance, we have a relation $-\ii \sigma_i^x \sigma_j^x w(a_x,a_y,a_z)=w(a_x+2,a_y,a_z)$, and similarly for other $a$'s. Rotating $w$ by $\pi/2$ around $y$-axis, $R_y=\exp{(-\ii \sigma^y \pi/4)}$, we get $R_y \otimes R_y w(a_x,a_y,a_z) R_y^\dagger \otimes R_y^\dagger=w(a_z,a_y,a_x)$. There is also a symmetry between $1+a_x$ and $1-a_x$, $\ii \sigma_i^x w^*(1+a_x,a_y,a_z)\sigma_j^x=w(1-a_x,a_y,a_z)$, as well as between positive and negative parameters, for instance, $\sigma_i^z w(a_x,a_y,a_z)\sigma_i^z=w^\dagger(a_x,a_y,-a_z)$. Becouse entanglement produced by $w^\dagger$ and $w^*$ is the same as by $w$, all these symmetries mean that it is enough to consider $a$'s in the following range:
\begin{equation}
1\ge a_x \ge a_y \ge a_z \ge 0.
\label{eq:a}
\end{equation}   
When doing numerics we will pay special attention to three choices of two qubit matrices. For gate $W$ we are going to consider a CNOT and an XY gate with the corresponding unitary matrices in standard computational basis equal to
\begin{equation}
W_{\rm CNOT}=
\begin{pmatrix}
1 & 0 & 0 & 0\\
0 & 1 & 0 & 0 \\
0 & 0 & 0 & 1\\
0 & 0 & 1 & 0
\end{pmatrix}
,\quad
W_{\rm XY}=
\begin{pmatrix}
1 & 0 & 0 & 0\\
0 & 0 & -i & 0 \\
0 & -i & 0 & 0\\
0 & 0 & 0 & 1
\end{pmatrix}.
\label{eq:W}
\end{equation}
The canonical form of CNOT has parameters $w_{\rm CNOT}(1,0,0)$ whereas XY has canonical form $w_{\rm XY}(1,1,0)$. Note that the number of nonzero $a$'s directly gives the number of CNOT gates needed to make such a gate out of single qubit gates and CNOT's\cite{Vidal:04}. In fact, XY gate is equivalent (has the same canonical form) as the product of two CNOTs, $W^{\rm CNOT}_{ij} W^{\rm CNOT}_{ji}$, also called a DCNOT gate. Third case will be a two qubit unitary $U_{ij}(t)$ chosen from U(4) invariant Haar measure~\footnote{As far as entanglement is considered this is the same as choosing $W$ in (\ref{eq:U}) as independent U(4) matrix at each step.}, shortly U(4) gate. Note that only in this case we choose an independent two qubit gate at each step, in all other cases two qubit gate is the same for all steps. This last choice of U(4) gate will serve for comparison. Naively, one might think that such random two qubit transformation at each step will produce entanglement in the fastest way. This is not so though. The optimal gate will turn out to be the XY gate.  
Analytical calculation of purity for general gate $w(a_x,a_y,a_z)$ and our protocol (\ref{eq:W}) is rather difficult. Things simplify though for gates $W$ which preserve the Pauli group.

\section{Special case : Markov chain}
\label{sec:markov}
Let us expand a pure state density matrix over products of Pauli matrices,  
\begin{equation}
\rho=\ket{\psi(t)}\bra{\psi(t)}=\sum_\mathbf{\alpha}{c_\mathbf{\alpha} \,\, \sigma_1^{\alpha_1} \cdots \sigma_n^{\alpha_n} },
\label{eq:c}
\end{equation}
where $\sigma_i^{\alpha_i}$ denotes Pauli matrix $\alpha_i \in \{0,x,y,z\}$ acting on $i$-th qubit, with the convention $\sigma^0=\mathbbm{1}$. We use a short notation $\mathbf{\alpha}=(\alpha_1,\ldots,\alpha_n)$. With the expansion (\ref{eq:c}) purity (\ref{eq:I}) is now simply given by 
\begin{equation}
I(t)=\frac{1}{N^2} \sum_{\mathbf{\alpha} =  \{\alpha_A 0_B \}} c_\alpha^2(t) ,
\label{eq:Ic}
\end{equation}
where the summation runs over all $\mathbf{\alpha}$ which have identity on the subspace B, i.e., $\alpha_j=0,\quad j=n/2+1,\ldots,n$. To obtain the decay of purity we therefore have to calculate the time dependence of $c_\alpha^2(t)$. Averaging over U(2) invariant Haar measure of single qubit matrices $V_i$, $V_j$ (\ref{eq:U}) we get the transformation law
\begin{equation}
c^2_\alpha(t+1)=\sum_{\beta,\gamma,\delta} v_{\beta,\delta} v_{\gamma,\delta} R_{\alpha_i,\delta_i} R_{\alpha_j,\delta_j} c_{\beta}(t) c_{\gamma}(t),
\label{eq:gen}
\end{equation}
where $v_{\alpha,\beta}$ is defined by $W_{ij} \sigma^\alpha W_{ij}^\dagger=\sum_\beta v_{\alpha,\beta} \sigma^\beta$ and $R$ is a $4\times4$ matrix obtained from averaging over random single qubit gates and is equal to
\begin{equation}
R=\begin{pmatrix}
1 & 0 & 0 & 0\\
0 & \frac{1}{3} & \frac{1}{3} & \frac{1}{3}\\
0 & \frac{1}{3} & \frac{1}{3} & \frac{1}{3}\\
0 & \frac{1}{3} & \frac{1}{3} & \frac{1}{3}
\end{pmatrix}.
\label{eq:R}
\end{equation} 
If $W_{ij}$ preserves the Pauli group, i.e., if it transforms products of Pauli matrices into a product of some other Pauli matrices, that is if $v_{\beta_i\beta_j,\gamma_i\gamma_j}$ is nonzero and equal to $\pm 1$ or $\pm \ii$ only if $\gamma_i=\beta'_i$ and $\gamma_j=\beta'_j$, the transformation (\ref{eq:gen}) can be simplified to
\begin{eqnarray}
c_\alpha^2(t+1)=\sum_\beta (M^{(2)}_{ij})_{\alpha,\beta}\, c_\beta^2(t),\\
\label{eq:M}
(M^{(2)}_{ij})_{\alpha,\beta}= R_{\alpha_i,\beta'_i} R_{\alpha_j,\beta'_j}.\nonumber
\end{eqnarray}
Markov matrix $M^{(2)}_{ij}$ of dimension $16\times 16$ involves only $i$-th and $j$-th qubits. To get the transformation on all $n$ qubits we have to average over all couplings, resulting in~\cite{Plenio} 
\begin{equation}
c^2(t+1)=M\, c^2(t),\qquad M=\frac{1}{L}\sum_{ij} M^{(2)}_{ij},
\label{eq:MM}
\end{equation}
if $L$ is the number of all couplings, i.e., number of allowed pairs of qubits $i$ and $j$. As mentioned, we will consider three different couplings, random $i$ and $j$, nnPBC and nnOBC. Let us illustrate how $\beta'$ are determined from $\beta$ (for instance, in eq.(\ref{eq:M})) on the example of a CNOT gate. All 16 different products of Pauli matrices $\sigma_i^{\beta_i} \sigma_j^{\beta_j}$ can be numbered by $x=\beta_j+4\cdot\beta_i$. Transformation of all 16 products can now be stated by giving the transformations of $x$'s. For instance, the transformation $U_{\rm CNOT} \sigma^x \sigma^y U^\dagger_{\rm CNOT}=\sigma^y \sigma^z$ can be simply stated by $x=6$ going to $x'=\beta_j'+4\cdot\beta_i'=11$. All 16 $x'$ for CNOT can be written in a vector $x'=(0,1,14,15,5,4,11,10,9,8,7,6,12,13,2,3)$, denoting the transformation $x=(0,1,2,\ldots,14,15) \to x'$. It turns out that XY gate also preserves the Pauli group. Transformed $x$'s for XY gate can be written in a vector $(0,11,7,12,14,5,9,2,13,6,10,1,3,8,4,15)$. Markovian description (\ref{eq:MM}) greatly simplifies the analysis and this formulation~\cite{Plenio} will be used to calculate the decay of purity. Purity is given by a sum of certain $c_\alpha^2$ (\ref{eq:Ic}) and therefore its asymptotic decay will be determined by the second largest eigenvalue of the matrix $M$. It can be shown that $M$ (\ref{eq:MM}) has two eigenvalues equal to $1$. One corresponds to the identity operator on all qubits, and the other to the ergodic density uniform on the rest of the space. A third eigenvalue, $\lambda_3=1-\Delta$, then determines the asymptotic decay of purity as $I(t)-I(\infty) \asymp (1-\Delta)^t$.

In addition to two qubit gates that preserve the Pauli group, Markov dynamics of the form $c^2(t+1)=M c^2(t)$ can also be written for the case when at each step we choose an independent U(4) gate. Averaging over U(4) group gives $M^{(2)}$ which preserves identity, $\sigma^0_i \sigma^0_j \to \sigma^0_i \sigma^0_j$, while it uniformly mixes all other 15 possible products $\sigma_i^{\alpha_i} \sigma_j^{\alpha_j}$. Matrix elements are therefore $[M^{(2)}(U4)]_{0,0}=1$ and $[M^{(2)}(U4)]_{x,x'}=1/15$ if $x,x' \in \{1,\ldots,15\}$. 

Most of two qubit gates $w(a_x,a_y,a_z)$ though do not preserve the Pauli group and therefore transformation of $c^2(t)$'s can not be written as a Markov process (\ref{eq:M}). In such cases we will have to use a direct numerical simulation of our protocol, averaging over many different realizations to obtain the decay of $I(t)$.

In the next section we are going to present numerical results on the convergence rates. Most of the time we are going to focus on purity. At the end of the section we will see that other quantities like the von Neumann entropy or the Schmidt coefficients give essentially the same information. Convergence rate will be obtained by several different methods. When looking for the optimal gate, a direct simulation of the protocol is used to calculate $I(t)$ at some fixed time for different gates, thereby identifying the optimal one. Central quantity of interest is the scaling of the convergence rate with $n$. In this respect we numerically calculate the gap of the corresponding Markov chain for different numbers of qubits $n$. We find that the gap in all cases scales as $\Delta \sim 1/n$. Finally, we use the obtained gaps $\Delta$ to predict the asymptotic decay of purity.

\section{Numerical results}
\subsection{Optimal gate}
First we want to find an optimal two qubit gate $W$ (\ref{eq:U}) which will result in the fastest possible decay of purity (\ref{eq:I}). We have already discussed that due to U(2) invariance of single qubit gates $V_i$ we can limit the study to two qubit gates in the canonical form (\ref{eq:w}) with the parameters $a_{x,y,z}$ in the range $1 \ge a_x \ge a_y \ge a_z \ge 0$ (\ref{eq:a}). Becouse Markov description for the transformation of $c^2$ is not possible for general parameters we have to resort to direct numerical calculation of purity. For each set of parameters we calculated the average purity $I(T)$ at some fixed time $T$, from which we then deduced the expected decay rate $\kappa$ of purity, assuming the dependence 
\begin{equation}
I(t)-I(\infty)=\exp{(-\kappa\, t/n)}.
\label{eq:kappa}
\end{equation}
Optimal gate is then the one with the largest $\kappa$. We always took $n=8$ qubits and averaged over $1000$ protocol realizations. Time $T$ at which we calculated $\kappa$ was $30$ for random $i-j$ coupling and $50$ for nnPBC and nnOBC couplings. Note that becouse in each case $n$ and $T$ were fixed we could instead of $\kappa$ simply use $I$ as a criterion to determine the optimal gate. The reason to use $\kappa$ is, as we will see later, that such theoretical form of purity is predicted in most cases of Markovian dynamics.    

\begin{figure}[h]
\centerline{\includegraphics[width=70mm,angle=-90]{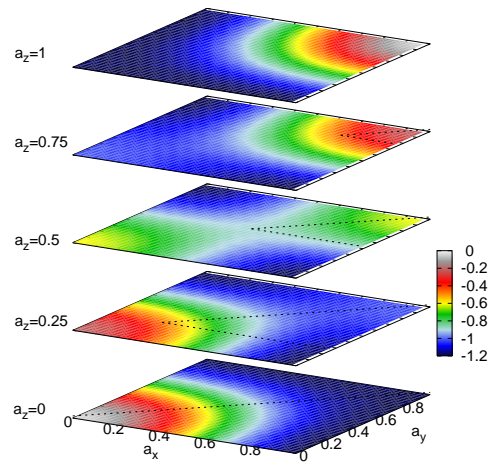}}
\caption{(Color online)~The dependence of the decay rate of purity $-\kappa$ (\ref{eq:kappa}) on three parameters $a_{x,y,z}$ for a protocol with random $i-j$ coupling. Cross-sections for four different $a_z$ are shown. Fastest decay ($\kappa \approx 1.2)$) is obtained for gates of the form $w(1,a_y,0)$. This includes XY gate at $w(1,1,0)$ as well as CNOT gate at $w(1,0,0)$. Dashed triangle shows the set of parameters fulfilling Eq.~(\ref{eq:a}).}
\label{fig:Iran}
\end{figure}

In Fig.~\ref{fig:Iran} we show the results in the case of coupling between random pairs of qubits. It seems that the fastest decay of purity is achieved for a continuous family of two qubit gates of the form $w(1,a_y,0)$ (and all variants obtained by symmetries). This includes XY gate with $w(1,1,0)$ as well as CNOT with $w(1,0,0)$. Gate $w(1,1,1)$, which is equivalent to a SWAP gate, does not produce any entanglement.  

\begin{figure}[h]
\centerline{\includegraphics[angle=-90,width=41mm]{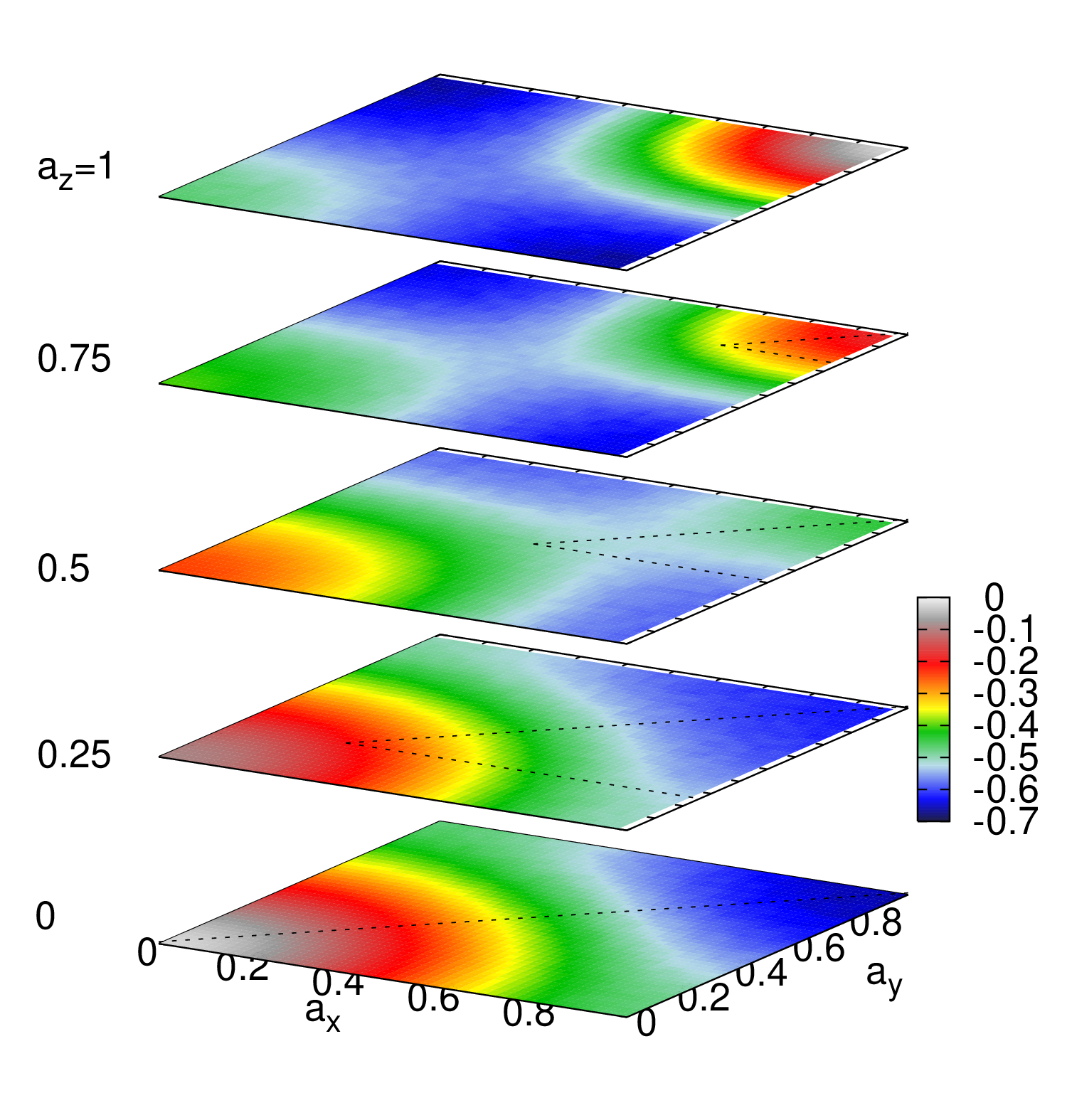}\includegraphics[angle=-90,width=41mm]{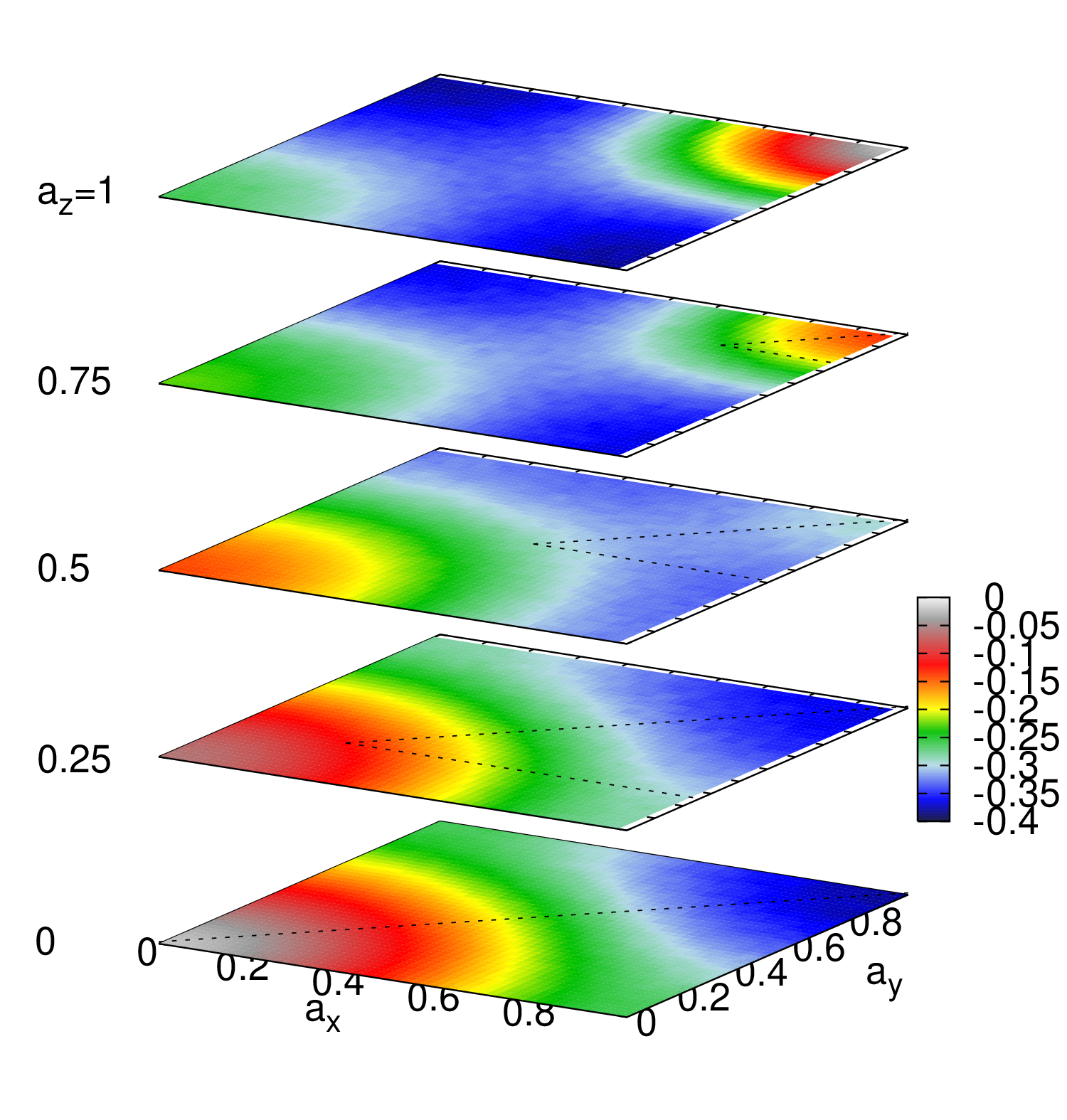}}
\caption{(Color online)~The dependence of the decay rate of purity $-\kappa$ (\ref{eq:kappa}) on three parameters $a_{x,y,z}$, similarly as in Fig.~\ref{fig:Iran}. On the left is for nnPBC coupling and on the right for nnOBC.}
\label{fig:Inn}
\end{figure}

In Fig.~\ref{fig:Inn} we show similar plots for nnPBC and nnOBC coupling. Apart from the numerical values of $\kappa$ the overall dependence on $a$'s looks similar to the one for random $i-j$ coupling. There is one notable difference thou. Now the XY and CNOT gate do not generate entanglement equally fast. For nnPBC for instance, $\kappa\approx 0.7$ for XY and $0.5$ for CNOT. XY is clearly better for generation of random entanglement. Similar conclusion is reached for nnOBC coupling. Interestingly, comparing nnPBC and nnOBC cases we can see that $\kappa$'s for the later are by a factor $\approx 2$ smaller than for nnPBC. Remember that the only difference between nnPBC and nnOBC is in the coupling between the first and the last qubit, i.e., in the term $U_{0,n-1}$, which is absent in nnOBC. This sole difference results in nnOBC being by a factor of $\approx 2$ slower in producing random entanglement. One can intuitively understand this factor in the following way. The average distance between qubits in subspaces $A$ and $B$ is in the case of nnPBC (qubits on a circle) about two times smaller than for nnOBC (qubits on a line). Therefore, becouse the decay rate for purity (\ref{eq:kappa}) scales inversely with $n$ one can expect a factor of $2$ between the two cases.    

\subsection{Markov chain}
\label{sec:MC}
In previous section we have identified two interesting gates, XY and CNOT, which are optimal for certain couplings. As we have seen in section~\ref{sec:markov} random protocol can be described by Markov chain for both of these gates. This has several advantages. In comparison to numerical calculation of purity averaging over protocol is in Markovian formulation exact. Provided the second largest eigenvalue (i.e., the largest smaller than $1$) of Markov matrix $M$ is nondegenerate, $1-\Delta$, the decay of purity will be for small gaps $\Delta$ given by 
\begin{equation}
I(t)-I(\infty)=\exp{(-t/\tau)},\qquad \tau=\frac{1}{\Delta}.
\label{eq:Itau}
\end{equation} 
Therefore, knowing the gap we will know the decay rate. Using analytical techniques one can actually bound the gap. This has been done in~\cite{Plenio} where they proved that the gap is $\Delta > \frac{4}{9n(n-1)}$ for CNOT gate and random $i-j$ coupling. As the exact analytical calculation of the gap seems too difficult, even for a relatively simple Markov chain with U(4) gate, we are going to numerically calculate the values of the gap for different $n$ and different couplings. In all cases we are going to consider XY, CNOT, and U(4) gate. Disadvantage of Markovian description is that the matrix $M$ is of rather large size $4^n$, i.e., their dimension equals to the square of the Hilbert space dimension of pure states.

\begin{figure}[h]
\centerline{\includegraphics[angle=-90,width=80mm]{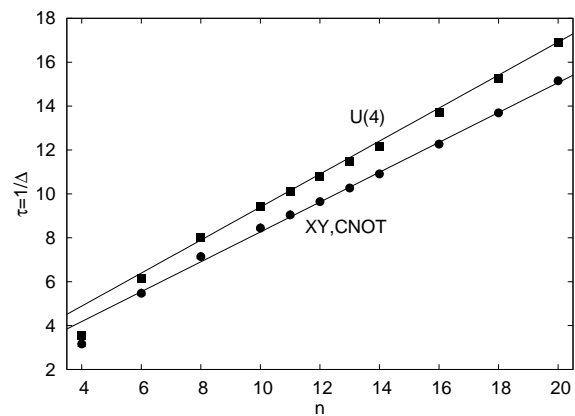}}
\caption{Dependence of the inverse gap, $1/\Delta$, on $n$ for Markovian chain with random $i-j$ coupling and XY (triangles), CNOT (circles) and U(4) (squares) gate. Full lines are best fitting linear function, see text for details.}
\label{fig:gapran}
\end{figure}

\begin{figure}[h]
\centerline{\includegraphics[angle=-90,width=80mm]{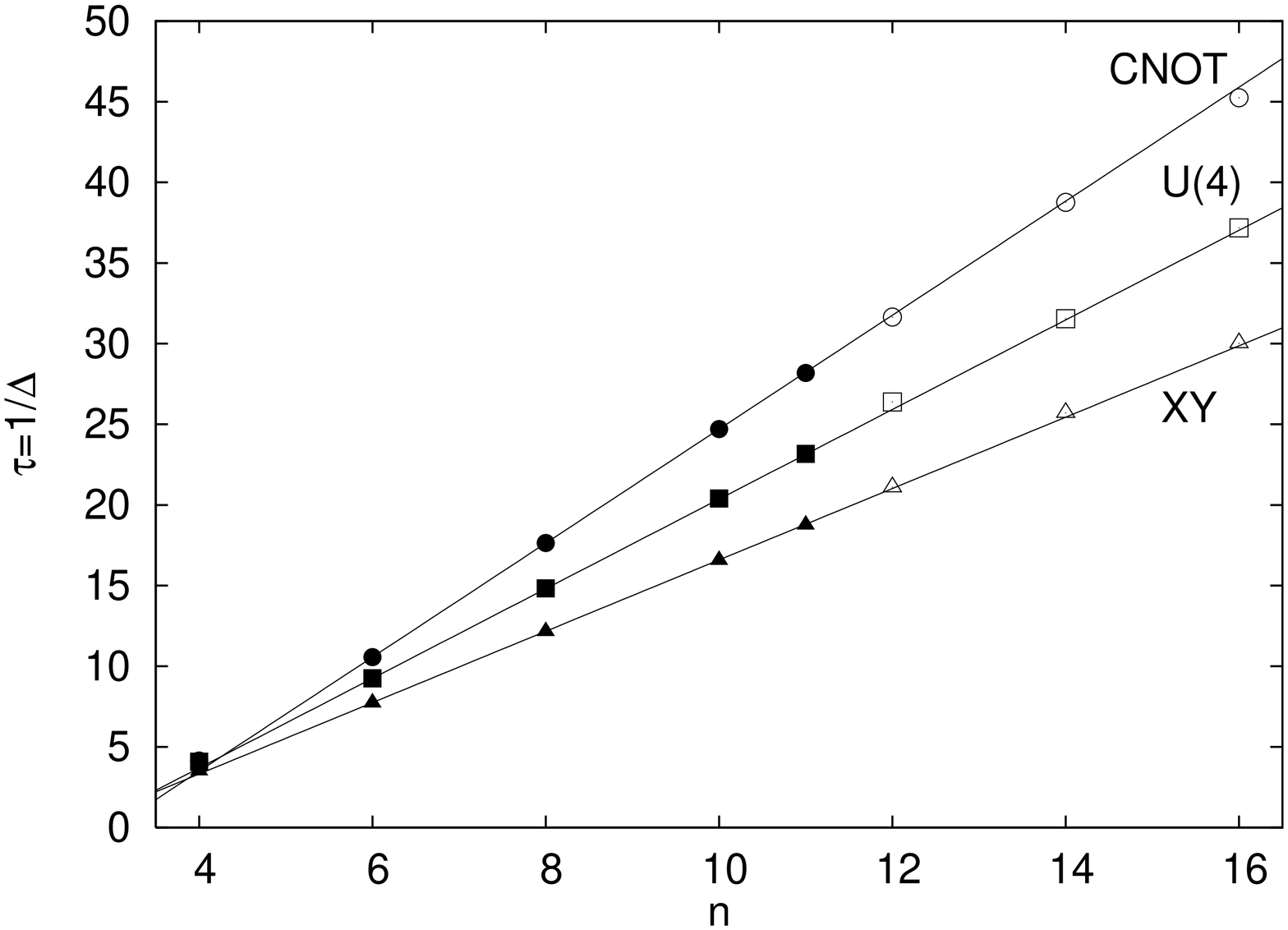}}
\caption{Dependence of the inverse gap $1/\Delta$ on $n$ for nnPBC coupling and XY (triangles), CNOT (circles) and U(4) (squares) gates. Full lines are best fitting linear functions, see text for details.}
\label{fig:gapnnPBC}
\end{figure}

Results of numerical calculation of the gap for a chain with random $i-j$ coupling are in Fig.~\ref{fig:gapran}. We use the Lanczos method~\cite{arpack} to calculate few largest eigenvalues of the Markovian matrix. For $n<13$ we use the original Markov chain $M$ of dimension $4^n$ while for larger $n$ a reduced chain is used whose size is only $2^n$, see~\cite{Plenio} for details. Becouse the size of the matrix grows exponentially with $n$ we are limited to relatively small number of qubits. In accordance with Fig.~\ref{fig:Iran} the gap is the same for XY and CNOT gates (same within numerical precision). Fitting $1/n$ dependence to the values of the gap we get $\Delta({\rm XY,ran-}ij)=1.47/(n+2.15)$ for XY and CNOT gate and $\Delta({\rm U4,ran-}ij)=1.33/(n+2.50)$ for U(4) gate~\footnote{Due to a small range of $n$'s available a logarithmic correction can not be excluded. In fact, numerical data are consistent with $1/\Delta({\rm U4,ran-}ij)=0.20n\ln{n}+4.7$ and $1/\Delta=0.18n\ln{n}+4.3$ for CNOT and XY gates. We would like to thank an anonymous referee for pointing out this possibility.}. Perhaps surprisingly, the U(4) is about 10\% slower than XY gate. Doing random two qubit gates at each step is therefore not as efficient in producing random entanglement as doing random single qubit gates and a fixed good two qubit gate (XY or CNOT)! Numerical result for the gap in the case of a CNOT gate improves the analytical lower bound $\Delta> 4/9n(n-1)$ proved in~\cite{Plenio}. 
For nnPBC results are shown in Fig.~\ref{fig:gapnnPBC}. Here full symbols correspond to numerical calculation of the gap on a $4^n$ dimensional Markov chain while empty symbols are gaps indirectly determined from the purity decay (\ref{eq:Itau}). The gap now scales as $\Delta({\rm XY,nnPBC})=0.45/(n-2.50)$, $\Delta({\rm U4,nnPBC})=0.36/(n-2.67)$ and $\Delta({\rm CNOT,nnPBC})=0.28/(n-3.01)$~\footnote{Here a logarithmic correction to the gap seems to be less likely.}. Here we see that XY is the optimal gate, while U(4) is better than CNOT gate, being the worst of the three. For nnOBC very similar results are obtained as for nnPBC and we will only list the fitted dependence of the gap, $\Delta({\rm XY,nnOBC})=0.23/(n-3.01), \Delta({\rm U4,nnOBC})=0.19/(n-3.12), \Delta({\rm CNOT,nnOBC})=0.15/(n-2.98)$.

\subsection{Purity decay}
While the asymptotic decay of purity is guaranteed to be $I(t)-I(\infty) \asymp \exp{(-t \Delta)}$ if the eigenvalue $1-\Delta$ of Markovian matrix is nondegenerate, the decay can be more complicated in the case of degeneracies. It is found that for many Markov chains one has the so-called cutoff phenomenon~\cite{Diaconis}. Briefly, we say there is a cutoff in a Markov chain if the distance between the asymptotic ergodic distribution and the distribution after $t$ steps sudenly drops from $1$ to $0$ at some cutoff time. The sharpness of this transition increases in the limit of large state space size (in our case for $n \to \infty$), for exact definition see~\cite{Diaconis}. Although precise mathematical conditions leading to the cutoff are not known~\cite{Diaconis}, it generally occurs due to multiplicity of the largest nontrivial eigenvalue with the degeneracy increasing with increasing $n$. Let us illustrate by an example: suppose we have $n$ times degenerate largest (non $1$) eigenvalue equal to $1-1/n$. Purity will then decay as $I \sim n\exp{(-t/n)}$. From this we see that $I$ will be equal to $\exp{(-c)}$, where $c$ is some fixed number, at time $t=n(\ln{n}+c)$. However large $c$ we choose, in the limit $n\to \infty$ purity will be small $I \sim \exp{(-c)}$ at time $n\ln{n}$. In other words, for times slightly smaller than $n\ln{n}$ purity will be very large while it will be very small for times slightly larger. There is a sudden cutoff at time $n\ln{n}$. On the other hand, if the eigenvalue is nondegenerate, purity is going to decay as $I \sim \exp{(-t/n)}$. It will reach small value $\exp{(-c)}$ at time $nc$. This time now increases with increasing $c$. There is no cutoff. If we want $I$ to be smaller we have to increase $t$. One can also look at time derivative of purity. In the case of a cutoff the absolute value of the derivative (steepness of the curve $I(t)$) diverges at some fixed value of $I$, while it goes to zero if there is no cutoff.  
\begin{figure}[h]
\centerline{\includegraphics[angle=-90,width=80mm]{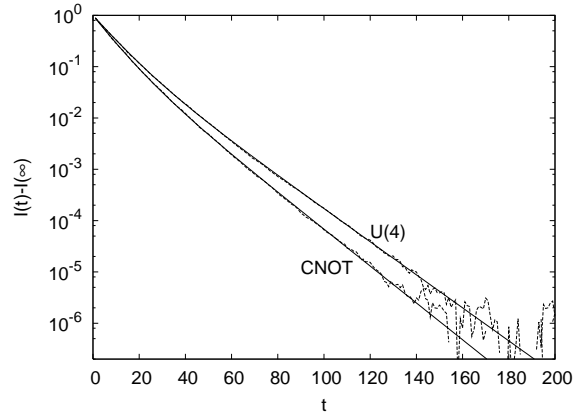}}
\caption{Decay of purity for $n=16$ and random $i-j$ coupling and U(4) and CNOT gates, dashed curves. Theoretical decay (\ref{eq:Ideg}) is shown with the full curve. Asymptotically purity decays as $\exp{(-t\Delta)}$ (\ref{eq:Itau}) with the gap predicted from Fig.~\ref{fig:gapran}. For smaller times thou degeneracies are important, reflected in a steeper decay of purity (\ref{eq:Ideg}). In the limit $n\to \infty$ the cutoff is predicted.}
\label{fig:I16ran}
\end{figure}

To identify a possible cutoff one therefore has to look at multiplicity of the largest eigenvalue. In all cases we studied in Sec.~\ref{sec:MC} the largest eigenvalue $1-\Delta$ is nondegenerate. However, large degeneracy of an eigenvalue that is very close to $1-\Delta$ could also possibly cause a cutoff. We have therefore numerically checked for degeneracies of the first three largest eigenvalues (here we mean first three nontrivial; we do not count a doubly degenerate eigenvalue $1$) for gates XY, CNOT, and U(4). In all three cases we found degeneracies only for random $i-j$ coupling, while there were no degenerate eigenvalues for nnPBC and nnOBC couplings. For XY and U(4) gate and random $i-j$ coupling a 3rd eigenvalues is $n-1$ times degenerate, while for CNOT gate a 2nd eigenvalues is $n-1$ times degenerate. This leads us to speculate that there is a cutoff for all three gates in the case of a random $i-j$ coupling~\cite{Plenio} while there is probably no cutoff for nnPBC or nnOBC coupling.    

To check the above prediction about the cutoff as well as to verify the asymptotic decay of purity, we have performed numerical calculation of purity decay for larger $n$, where the cutoff phenomenon should be visible. We simulated our random protocol, averaging over many realizations, thereby obtaining $I(t)$. In Fig.~\ref{fig:I16ran} we show the decay of purity for $n=16$ qubits and random $i-j$ coupling. As we have seen, the largest eigenvalue is in all cases nondegenerate while the 2nd or 3rd eigenvalue is $n-1$ times degenerate for random $i-j$ coupling. Therefore, for short times the decay of purity will be given by the largest as well as by the mentioned degenerate eigenvalue. We fitted numerically obtained curves in Fig.~\ref{fig:I16ran} with the following dependence
\begin{equation}
I(t)-I(\infty)=\frac{\left(1+a\cdot\exp{(-b\cdot t/\tau)}\right)}{1+a}\,\exp{(-t/\tau)},
\label{eq:Ideg}
\end{equation}
where $\tau=1/\Delta$ is given by the gap of the largest eigenvalue (\ref{eq:Itau}) and is determined from linear fitting lines in Fig.~\ref{fig:gapran}, while two fitting parameters $a$ and $b$ take care of the degeneracy and the gap of the largest degenerate eigenvalue, respectively. For the CNOT gate and $n=16$ we get $a=2.80$ and $b=0.75$, while for U(4) gate we get $a=2.57$ and $b=0.75$. Decay times $\tau$ (\ref{eq:Ideg}) used were the ones predicted from Fig.~\ref{fig:gapran} and are $\tau=12.3$ for CNOT and $\tau=13.9$ for U(4) gate. Similar fitting parameters as for CNOT gate are obtained also for XY gate (data not shown). Important point is that the parameter $a$, connected with the degeneracy, increases with $n$. For instance, for U(4) gate it is $a=0.78$ for $n=12$, $a=2.57$ for $n=16$ and $a=6.4$ for $n=18$. This signals the emergence of a cutoff for random $i-j$ coupling and large $n$.

\begin{figure}[h]
\centerline{\includegraphics[angle=-90,width=80mm]{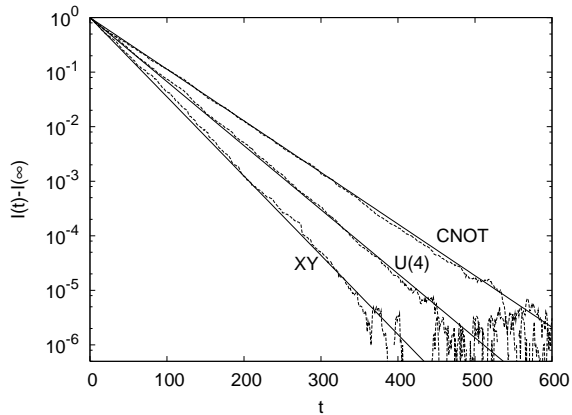}}
\caption{Same as Fig.~\ref{fig:I16ran} but for nnPBC coupling. Purity is well described by theoretical $I(t)-I(\infty)=\exp{(-t/\tau)}$ (\ref{eq:Itau}), with $\tau$ predicted in Fig.~\ref{fig:gapnnPBC} (i.e., no fitting), in the whole range of times. No cutoff is expected.}
\label{fig:I16nnPBC}
\end{figure}
On the other hand, things are quite different for nnPBC or nnOBC coupling. Results of numerical simulation for nnPBC are in Fig.~\ref{fig:I16nnPBC} (for nnOBC data are very similar). As one can see, purity is well described by a single exponential function (\ref{eq:Itau}) with the predicted $\tau=29.9$ for XY, $\tau=37.0$ for U(4) and $\tau=45.9$ for CNOT gate. Therefore, according to numerical results for sizes upto $n=20$, we can predict that there is probably no cutoff phenomenon for nnPBC and nnOBC couplings. 

Numerical results presented show that the decay time of purity asymptotically scales as $\tau \sim n$. This holds also for gates that can not be described by Markovian chain (data not shown). Becouse random state value of purity is $\sim 1/N$, this implies that purity will decay to its asymptotic value $I(t)-I(\infty)\sim 1/2^n$ after number of steps scaling as $t \sim n^2$.

\subsection{Measures in addition to purity}
\begin{figure}[h]
\centerline{\includegraphics[angle=-90,width=80mm]{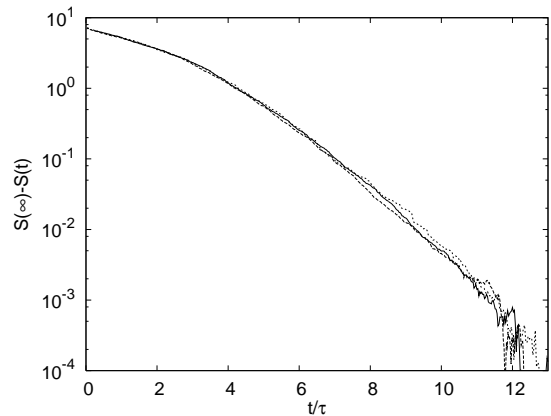}}
\caption{Similarly as $I(t)$, von Neumann entropy (\ref{eq:S}) also decays on the same time scale as purity. All is for nnPBC, gates XY, CNOT and U(4) and $n=16$.}
\label{fig:S}
\end{figure}
In the core of the paper, due to its simplicity, we have used purity $I(t)$ as a measure of entanglement. Here we are going to show that other quantities, von Neumann entropy and Schmidt coefficients, for instance, also decay on the same time scale $\tau=1/\Delta$ as purity. The only difference is that the functional dependence of the decay is more complicated than a simple exponential function. First we show the results for the von Neumann entropy $S(t)$ (\ref{eq:S}). In order to show the scaling with decay time $\tau$ we plot in Fig.~\ref{fig:S} the dependence of $S$ on $t/\tau$. As we can see, decay of $S(t)$ is indeed described by a universal form $S(\infty)-S(t)=f(t/\tau)$, with the function $f$ visible in figure. It slightly depends on the type of the coupling, i.e., whether we have a random $i-j$, nnPBC or nnOBC, but is independent of the gate.

\begin{figure}[h]
\centerline{\includegraphics[angle=-90,width=43mm]{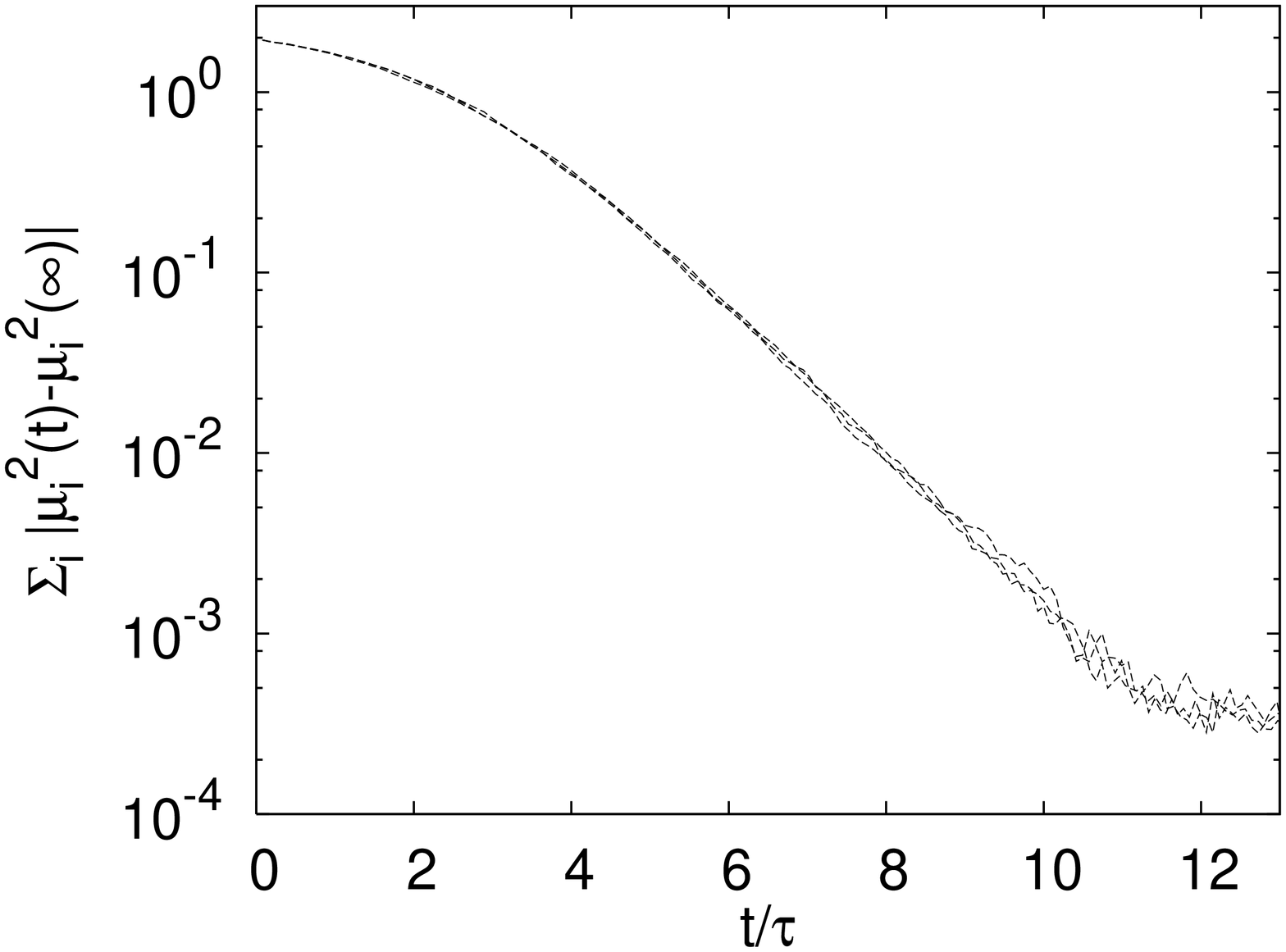}\includegraphics[angle=-90,width=43mm]{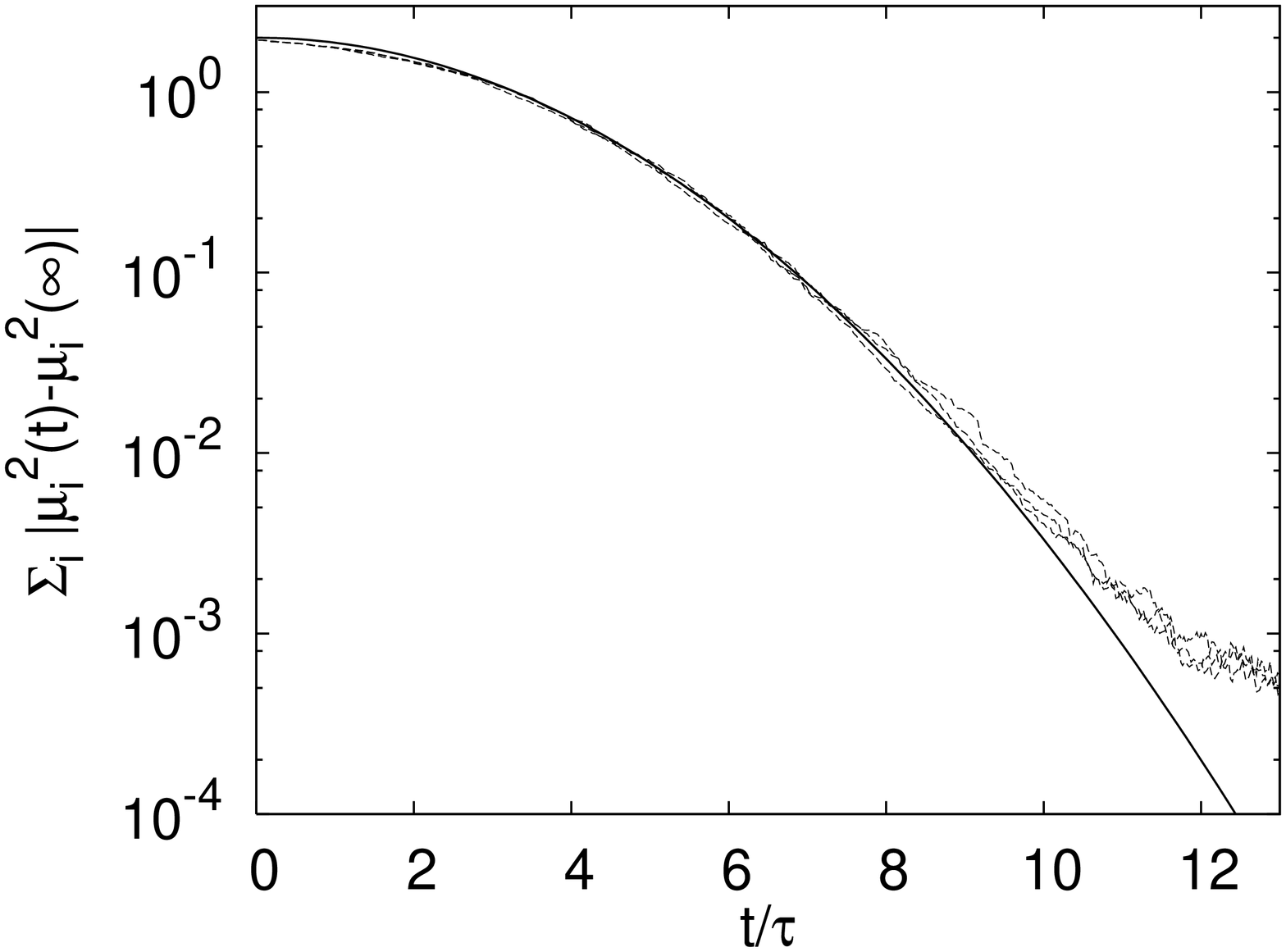}}
\caption{Schmidt coefficients also converge on the same time scale $\tau$ as purity. We show how $\sum_i|\mu_i^2(t)-\mu_i^2(\infty)|$ depends on $t/\tau$. On the left is for random $i-j$, on the right for nnPBC coupling (full line is a Gaussian fit). In both plots three curves almost overlapping are for CNOT, XY, and U(4) gates, $n=16$.}
\label{fig:Sch}
\end{figure}
\begin{figure}[h]
\centerline{\includegraphics[angle=-90,width=80mm]{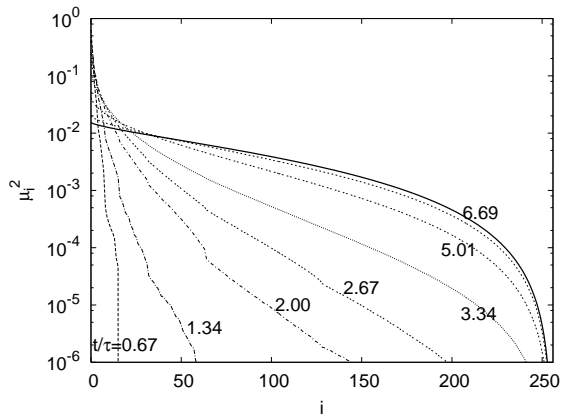}}
\caption{Convergence of squares of Schmidt coefficients $\mu_i^2(t)$ (i.e., eigenvalues of the reduced density matrix) for $n=16$ and XY gate with nnPBC. Theoretical decay time is $\tau=29.9$. Full thick line is theoretical prediction $\mu_i^2(\infty)$ (\ref{eq:mui}).}
\label{fig:mui}
\end{figure}

To fully specify bipartite entanglement a single quantity, like purity or von Neumann entropy, is not enough. One has to specify all Schmidt coefficients $\mu_i$. As a final test we checked how individual Schmidt coefficients converge to those of random pure states. Theoretical prediction for random pure states has been calculated in~\cite{JPA} and is given in an implicit form by Eq.~(\ref{eq:mui}). Again, Fig.~\ref{fig:Sch} shows the convergence on the same time scale as that of purity. Functional dependence looks Gaussian for nnPBC coupling while it is more complicated for random $i-j$ coupling, possibly due to degeneracies. In Fig.~\ref{fig:mui} we show the convergence of individual $\mu_i^2$.

\section{Discussion}
We have numerically studied protocols for generation of generic entanglement as represented by random pure quantum states. At each step of the protocol a fixed two qubit gate and two independent random single qubit unitaries are applied. We have calculated the decay rate of purity which in all cases studied grows linearly with the number of qubits $n$. For certain two-qubit gates Markovian description is possible. Convergence rate is in such cases determined by the size of the gap which is numerically found to scale as $1/n$, improving the analytical bound in~\cite{Plenio}. An optimal two qubit gate is identified which produces random entanglement in the smallest number of steps. This optimal gate is for all different couplings considered XY gate, also called DCNOT gate, generated by the Heisenberg XX interaction. Depending on the coupling XY gate can be as much as 60\% faster than CNOT gate. Interestingly, applying a random two qubit gate at each time step is slower than applying a fixed good two qubit gate in addition to random single qubit unitaries. For coupling between random qubits we predict the cutoff phenomenon, while there is probably no cutoff for the protocol where we apply two qubit gates only to nearest neighbors.

The author is indebted to an anonymous referee for valuable suggestions. Financial support by Slovenian Research Agency, programme P1-0044, and grant J1-7437, is gratefully acknowledged.

\end{document}